\documentclass[twocolumn,showpacs,preprintnumbers,amsmath,amssymb,prl,aps]{revtex4}

\usepackage{epsfig}
\usepackage{graphicx}
\usepackage{dcolumn}
\usepackage{bm}

\begin{document}

\title{Evidence for strong 5$d$ electron correlations and electron-magnon coupling
in a pyrochlore, Y$_2$Ir$_2$O$_7$}

\author{R.S. Singh}
\author{V.R.R. Medicherla}
\author{Kalobaran Maiti}
\altaffiliation{Electronic mail: kbmaiti@tifr.res.in}
\author{E.V. Sampathkumaran}

\affiliation{Department of Condensed Matter Physics and Materials
Science, Tata Institute of Fundamental Research, Homi Bhabha Road,
Colaba, Mumbai - 400 005, INDIA}

\date{\today}

\begin{abstract}

We report the observation of an unusual behavior of highly extended
5$d$ electrons in Y$_2$Ir$_2$O$_7$ belonging to pyrochlore family of
great current interest using high resolution photoemission
spectroscopy. The experimental bulk spectra reveal an intense lower
Hubbard band in addition to weak intensities in the vicinity of the
Fermi level, $\epsilon_F$. This provides a direct evidence for
strong electron correlation among the 5$d$ electrons, despite their
highly extended nature. The high resolution spectrum at room
temperature exhibits a pseudogap at $\epsilon_F$ and $|\epsilon -
\epsilon_F|^2$ dependence demonstrating the importance of electron
correlation in this system. Remarkably, in the magnetically ordered
phase ($T <$~150~K), the spectral lineshape evolves to a $|\epsilon
- \epsilon_F|^{1.5}$ dependence emphasizing the dominant role of
electron-magnon coupling.

\end{abstract}


\maketitle

It is generally believed that electron-electron Coulomb repulsion
strength, $U$, among the $d$ electrons decreases as one traverses
from 3$d$ to 5$d$ ions, due to the increase in radial extension of
the these orbitals. For example, while the effective electron
correlation strength, $U/W$ ($W$ = bandwidth) in 3$d$ transition
metal oxides (TMOs) is significantly strong \cite{fujimori-review},
$U/W$ in 4$d$ TMOs is weak enabling first principle approaches to
capture the experimentally determined electronic structure
\cite{ruthPRBR,fujimori}. Signature of electron correlation in 5$d$
transition metal compounds has not been observed
\cite{allen,Cd2Re2O7-2,bairo3band}. In this letter, we offer a clear
evidence for strong correlation among 5$d$ electrons in a compound,
Y$_2$Ir$_2$O$_7$, which has not been paid enough attention in the
literature.

This compound belongs to pyrochlore structure, which have drawn
significant attention in recent times due to the possibility of
geometrical frustration leading to a varieties of novel phenomena
e.g., spin ice behavior \cite{SpinIce}, superconductivity
\cite{Cd2Re2O7-1}, correlation induced metal insulator transitions
\cite{R2Mo2O7} {\em etc}. In particular, a Ir based pyrochlore,
Pr$_2$Ir$_2$O$_7$ has been recently identified to show spin-liquid
behavior \cite{SLpr2ir2o7}, anomalous Hall effect \cite{AHEpr2ir2o7}
{\em etc}. It is believed that the interaction between the Pr 4$f$
moments mediated by Ir 5$d$ conduction electrons plays a key role
here. While Pr$_2$Ir$_2$O$_7$ is a metal, the Y-analogue,
Y$_2$Ir$_2$O$_7$ is an insulator. Y$_2$Ir$_2$O$_7$ is described to
be a {\em Mott insulator}, where large $U/W$ leads to insulating
phase in a metal \cite{Fukazawa,soda,fukazawa1}. In addition,
Y$_2$Ir$_2$O$_7$ has been proposed to exhibit a weak ferromagnetic
transition at around 150~K \cite{soda,fukazawa1}. Therefore, this
system appears to be an ideal system to probe 5$d$ electron
correlation.

In this letter, we report the results of our investigation on the
origin of unusual properties in Y$_2$Ir$_2$O$_7$ using
state-of-the-art high resolution photoemission spectroscopy. We find
that the bulk Ir 5$d$ band exhibits a signature of intense lower
Hubbard band as observed in strongly correlated 3$d$ TMOs
\cite{fujimori-review,kbmPRL}. It is remarkable that no hard gap is
observed in the bulk spectra although it is an insulator. The
spectral lineshape in the vicinity of the Fermi level, $\epsilon_F$
changes from $|\epsilon - \epsilon_F|^2$ behavior at 300 K
corresponding to a correlated Fermi liquid system to $|\epsilon -
\epsilon_F|^{1.5}$ behavior in the ferromagnetic phase suggesting
importance of electron-magnon coupling.

High quality Y$_2$Ir$_2$O$_7$ was prepared by solid state reaction
route using high-purity ($>$~99.9\%) ingredients (Y$_2$O$_3$ and
IrO$_2$ powders). To achieve large grain size and good intergrain
binding, the sample was sintered in pellet form at 900~$^o$C for a
day and subsequently, at 1000~$^\circ$C for more than two days with
an intermittent grinding. The sample quality was characterized by
x-ray diffraction and scanning electron microscopy. There was no
evidence for any impurity feature. Rietveld refinement reveals a
single cubic phase ($a$~=~10.20~\AA; space group $Fd{\bar{3}}m$).
The sample was further characterized \cite{kartik} by magnetization
measurements and found to show a magnetic transition at about 150~K
in agreement with the literature \cite{soda,fukazawa1}.
Photoemission measurements in the temperature range 10 K - 300 K
were performed using monochromatic photon sources and SES2002
Gammadata Scienta analyzer. The energy resolutions were set to
300~meV, 4~meV and 1.4~meV for the measurements with Al $K\alpha$
(1486.6~eV), He {\scriptsize II} (40.8~eV) and He {\scriptsize I}
(21.2~eV) photons, respectively. The sample surface was cleaned by
{\it in situ} scraping (base pressure = 3$\times$10$^{-11}$~torr)
and cleanliness of the sample surface was ensured by negligible
($<$~2\%) impurity contributions in the O $1s$ spectral region and
the absence of C $1s$ peak. The electronic band structure
calculations were carried out using full potential linearized
augmented plane wave method (WIEN2k software)\cite{wien} within the
local density approximations, LDA. The convergence was achieved
considering 512 $k$ points within the first Brillouin zone and the
error bar for the energy convergence was set to $\sim$~0.25~meV.

In Fig. 1, we show the density of states (DOS) calculated for the
nonmagnetic ground state representing the electronic structure at
room temperature. The total density of states (TDOS), Ir 5$d$
partial density of states (PDOS), O1 2$p$ PDOS and O2 2$p$ PDOS are
plotted in different panels of the figure. Here, O1 and O2 represent
the oxygens forming IrO1$_6$ octahedra and the oxygens bonded to Y
forming O2Y$_4$ tetrahedra, respectively. Y 4$d$ PDOS appear
primarily beyond 4.7~eV above $\epsilon_F$ with negligible
contribution in the occupied part and hence not shown here. There
are two groups of features in the TDOS in the occupied part. The
energy range, -7.5 to -1.5~eV is primarily dominated by O1 2$p$ and
O2 2$p$ PDOS with a small contribution from Ir 5$d$ PDOS. O2 2$p$
PDOS appears between -5.0 to -1.5 eV, which is not bonded to Ir 5$d$
states. O1 2$p$ - Ir 5$d$ bonding states with dominant O1 2$p$
character contribute to the energy range below -5 eV. In the energy
range -1.2~eV to 0.3~eV, the antibonding bands having primarily Ir
5$d$ character with $t_{2g}$ symmetry appear. It is to be noted that
there is a large TDOS at $\epsilon_F$, suggesting a metallic phase
in contrast to the insulating behavior observed in transport data
\cite{Fukazawa,soda}.

\begin{figure}
\vspace{-2ex}
\includegraphics[angle=0,width=0.4\textwidth]{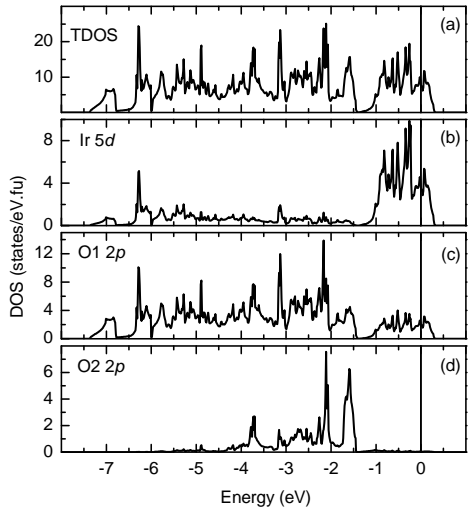}
\vspace{-12ex}
\caption{ Calculated (a) Total density of states (TDOS), (b) Ir 5$d$
partial density of states (PDOS), (c) O1 2$p$ PDOS and (d) O2 2$p$
PDOS. Zero in the energy axis represents $\epsilon_F$.}
 \vspace{-2ex}
\end{figure}

\begin{figure}
\vspace{-2ex}
\includegraphics[angle=0,width=0.38\textwidth]{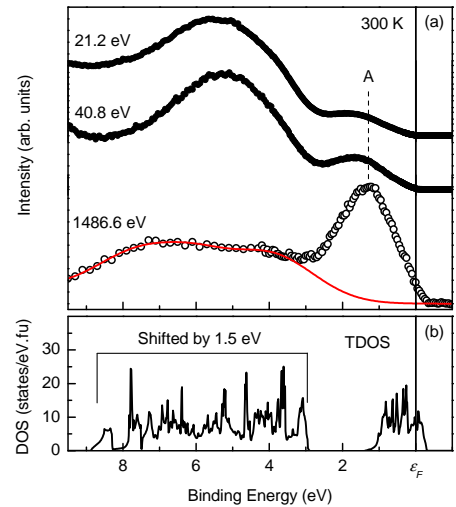}
\vspace{-12ex}
\caption{(color online) (a) Valence band spectra collected using
photon energies 21.2~eV, 40.8~eV and 1486.6~eV. (b) Calculated TDOS
shown in Fig.~1. The O 2$p$ part is shifted to match the
experimental spectra.}
 \vspace{-2ex}
\end{figure}

The experimental valence band photoemission spectra at room
temperature are shown in Fig.~2(a). The spectra corresponding to
ultraviolet (UV) photons (21.2 eV and 40.8 eV) exhibit two
distinctly separated features. A dominant intensity appears between
3 - 9 eV binding energies (= $\epsilon_F$ - $E$, $E$ = energy) and a
weak feature, A, near $\epsilon_F$. The intensity pattern becomes
drastically opposite in the 1486.6 eV spectrum; the feature A
becomes the most intense one compared to the intensity of the other
feature. The ratio of the photoemission cross section of Ir 5$d$
states to the O 2$p$ states increases significantly for 1486.6 eV
incident photon energy compared to UV energies; the feature A in the
figure can therefore be attributed to the photoemission signal from
the bands having essentially Ir 5$d$ character. The O 2$p$ features
appear above 3~eV binding energy. A rigid shift of the calculated O
2$p$ band by about 1.5 eV towards higher binding energies provides a
remarkable representation of the experimental spectra as shown in
Fig.~2(b). Such a rigid shift of the completely filled O 2$p$ bands
is often observed due to the underestimation of the correlation
effects in the band structure calculations \cite{ddPRL}.

It is clear from Figs.~1 and 2 that O 2$p$ and Ir 5$d$ related
features are distinctly separated. Thus, Ir 5$d$ contributions
appearing near $\epsilon_F$ can be delineated by subtracting the
tail of the O 2$p$ band. We have simulated the O 2$p$ contributions
using a combination of Lorentzians convoluted by a Gaussian
representing resolution broadening as shown by solid line in the
case of 1486.6 eV spectrum in Fig.~2(a). The extracted Ir 5$d$ bands
are shown in Fig.~3(a) after normalizing to the integrated
intensity. The 1486.6~eV spectrum clearly shows different lineshape
and energy position compared to the UV spectra. This is demonstrated
by overlapping the resolution broadened 21.2~eV spectrum (dashed
line) over the 1486.6~eV spectrum. The $x$-ray photoelectrons have
larger escape depth than the ultraviolet photoelectrons. Thus, the
difference in the 1486.6 eV and UV spectra is attributed to the
different bulk and surface electronic structures.

\begin{figure}
\vspace{-2ex}
\includegraphics[angle=0,width=0.38\textwidth]{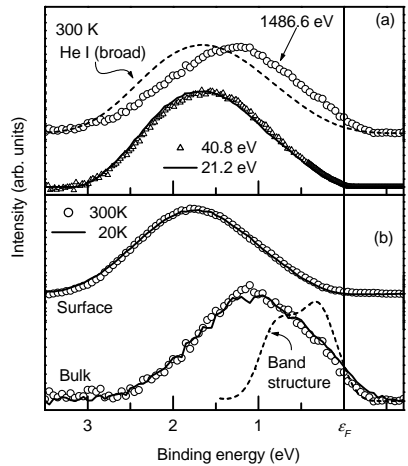}
\vspace{-16ex}
\caption{(color online) (a) Ir 5$d$ band at different photon
energies. Dashed line represents the resolution broadened 21.2 eV
spectrum (solid line). (b) Extracted surface and bulk spectra at 300
K (symbols) and 20 K (solid line). Dashed line is the calculated
spectral function from band structure results.}
 \vspace{-2ex}
\end{figure}

The intensity in the photoemission spectra can be expressed as
$I(\epsilon) = (1-e^{-d/\lambda}).I^s(\epsilon) + e^{-d/\lambda}.
I^b(\epsilon)$, where $I^s(\epsilon)$ and $I^b(\epsilon)$ are the
surface and bulk spectra, respectively. $d$ is the effective surface
layer depth and $\lambda$ is escape depth of the photoelectrons. To
calculate $I^s(\epsilon)$ and $I^b(\epsilon)$, we use $d/\lambda$ =
1.8 and 0.5 for UV and $x$-ray photoemission spectra, respectively
as used for several other systems \cite{ruthPRBR,kbmPRB}. The
extracted surface and bulk spectral functions are plotted in Fig.
3(b). The surface spectrum exhibits a peak around 1.8~eV with no
intensity at $\epsilon_F$ suggesting insulating character of the
surface electronic structure. The bulk spectrum, on the other hand,
exhibits a substantial intensity at $\epsilon_F$ suggesting metallic
phase in addition to an intense peak at around 1~eV.

In order to address the issue of electron correlation among 5$d$
electrons, we compare the surface and bulk spectra with the {\em ab
initio} results. The dashed line in Fig.~3(b) represents the
Gaussian-convoluted (full width at half maximum = 0.3 eV) occupied
TDOS at 300~K. The calculated spectrum exhibits a peak at around 0.5
eV, spreading down to about 1 eV binding energy, and it is
significantly narrower than the width of the bulk spectra. Since,
the electron correlation is significantly underestimated in the band
structure calculations within LDA, the difference between the
calculated spectrum and the experimental one is often attributed to
the electron correlation effects \cite{fujimori-review}. Thus, the
intensities appearing at higher binding energies in the bulk
spectrum represent photoemission signal from the correlation induced
localized electronic states (lower Hubbard band), and is termed as
'incoherent feature'. The intensities in the vicinity of
$\epsilon_F$ represent the signature of delocalized electronic
states and is termed as 'coherent feature'. The large intensity of
the incoherent feature compared to the coherent feature intensity
indicates strong electron correlation effects, in sharp contrast to
the observations in other Ir compounds \cite{bairo3band}. Such a
strong correlation in the highly extended 5$d$ bands is unusual. The
deviation from LDA results is most evident in the surface spectrum.
Only the incoherent feature is observed along with a large gap at
$\epsilon_F$ suggesting a Mott insulating phase corresponding to the
two dimensional electronic structure at the surface.

We now focus on the influence of the magnetic phase transition on
the electronic structure.

For this purpose, we first discuss the surface and bulk spectra at
20~K shown by solid lines in Fig.~3(b). The surface spectra remain
unaffected with the change in temperature. No hard gap is observed
in the bulk spectra down to the lowest temperature studied, which
evidently rules out the possibility of Mott insulating phase in the
bulk even at low temperatures. The lineshape of the 20~K bulk
spectrum is very similar to that at 300~K indicating that the
magnetic phase transition has insignificant influence if viewed in
the energy scale of the figure. However, a closer look near
$\epsilon_F$ region reveals subtle changes with temperature as
described below.

Although the electron correlation effects are manifested in the
large energy scale as described above, various thermodynamic
properties are essentially determined by the electronic states near
$\epsilon_F$ ($|\epsilon - \epsilon_F| \approx k_BT$). High energy
resolution employed in the present investigation enables to address
this issue. We investigate the evolution of the He~{\scriptsize I}
spectra near $\epsilon_F$ as a function of temperature in Fig.~4(a),
which represent the bulk features as the surface spectra exhibit a
large gap (see Fig.~3(b)). Normalization of all the spectra at
around 200~meV binding energy shows similar line shape down to about
50 meV binding energy at all the temperatures. The spectra in the
energy range closer to $\epsilon_F$ reveal interesting evolution
along with the appearance of a sharp Fermi cut off at low
temperatures.

\begin{figure}
\vspace{-4ex}
\begin{center}
\includegraphics[angle=0,width=0.4\textwidth]{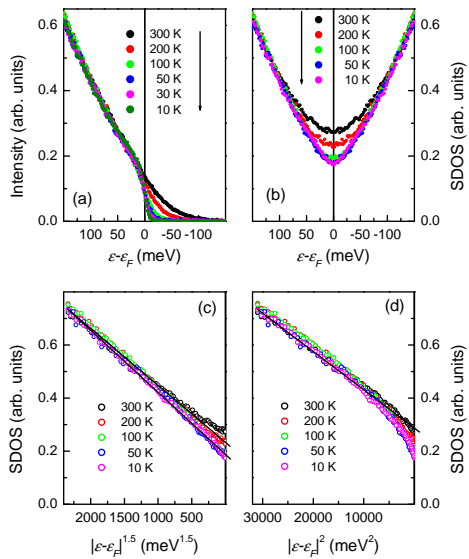}
\vspace{-12ex}
\end{center}
\caption{(color online) (a) High resolution spectra near
$\epsilon_F$ as a function of temperature. (b) SDOS obtained by
symmetrizing the spectra in (a). SDOS is plotted as a function of
(c) $|\epsilon-\epsilon_F|^{1.5}$ and (d) $|\epsilon-\epsilon_F|^2$.
Zero in the energy axis denotes $\epsilon_F$.}
 \vspace{-2ex}
\end{figure}

Since the energy resolution is very high and various lifetime
broadenings are insignificant in the vicinity of $\epsilon_F$, one
can extract the spectral density of states (SDOS) directly from the
raw data by symmetrizing the spectra; SDOS~=~$I(\epsilon) +
I(-\epsilon)$. Thus obtained SDOS shown in Fig.~4(b), provide a good
representation of the density of states in this small energy window.
There is a dip in SDOS at $\epsilon_F$, which gradually increases
with the decrease in temperature down to 100~K. Further reduction in
temperature does not show any significant change in the intensity at
$\epsilon_F$. It is often observed that the intensity at
$\epsilon_F$ decreases with the decrease in temperature due to the
disorder induced localization of the electronic states at
$\epsilon_F$ \cite{DDdisorder,kobayashi}. In such a case, the DOS at
$\epsilon_F$ follows ($a + b\sqrt{T}$; $T$ is temperature) behavior
\cite{kobayashi,altshuler-aronov}, which is qualitatively different
from the behavior in the present case.

In order to investigate the energy dependence of the spectral
lineshape, we analyzed SDOS as a function of $|\epsilon -
\epsilon_F|^\alpha$ for different values of $\alpha$. All the
spectra could not be simulated for one value of $\alpha$. We show
the two extreme cases in Fig.~4(c) and 4(d). The SDOS at 300~K is
better represented by $\alpha$~=~2 indicating strong influence of
electron correlation effect in the electronic structure. This
finding may serve as an experimental demonstration of the
theoretical prediction of correlation induced effects in a Fermi
liquid \cite{Correlation}. Clearly, disorder does not play a major
role as corresponding behavior of $\alpha$ = 0.5 is not observed
\cite{DDdisorder,kobayashi,altshuler-aronov}. The spectra in the
magnetically ordered phase ($T <$~150~K) exhibit a deviation from
$\alpha$~=~2 behavior and are better represented by $\alpha$~=~1.5.
High resolution spectra of another 5$d$ compound, BaIrO$_3$,
exhibiting ferromagnetic ground state below 183~K is also
characterized by $\alpha$~=~1.5 \cite{bairo3}. Thus, it is clear
that electron-magnon coupling \cite{irkhin} plays a significant role
to determine the magnetically ordered states in these highly
extended 5$d$ systems.

In summary, we have addressed the issue of electron correlation
within 5$d$ band by choosing Y$_2$Ir$_2$O$_7$, a compound belonging
to a family of great current interest. A comparison of the
electronic structure calculations and the high-resolution
photoemission data offers distinct evidence for strong Coulomb
correlation among Ir 5$d$ electrons. No hard gap is observed in the
bulk spectra down to the lowest temperature studied. The surface
spectra exhibit an insulating phase at all the temperatures
presumably due to the enhancement of $U/W$ at the surface compared
to bulk. Analysis of the spectral lineshape in the vicinity of
$\epsilon_F$ as a function of temperature emphasizes the need to
consider electron-magnon coupling in addition to electron
correlation while trying to understand the solid state behavior of
the 5$d$ electrons.

We thank K.K. Iyer for his valuable help in preparation and
characterization of the sample.


\begin{thebibliography}{99}
%
\bibitem{fujimori-review} M. Imada, A. Fujimori, and Y. Tokura, Rev.
Mod. Phys. {\bf 70}, 1039 (1998).
%
\bibitem{ruthPRBR} K. Maiti and R.S. Singh, Phys. Rev. B {\bf 71},
161102(R) (2005).
%
\bibitem{fujimori} M. Takizawa {\em et al.}, Phys. Rev. B {\bf 72},
060404(R) (2005).
%
\bibitem{allen} Li-Shing Hsu, G. Y. Guo, J. D. Denlinger, and J. W.
Allen, Phys. Rev. B {\bf 63}, 155105 (2001).
%
\bibitem{Cd2Re2O7-2} R. Eguchi {\em et al.}, Phys. Rev. B {\bf
66}, 012516 (2002).
%
\bibitem{bairo3band} Kalobaran Maiti, Phys. Rev. B {\bf 73},
115119 (2006).
%
\bibitem{SpinIce} S. T. Bramwell and M. J. P. Gingras, Science {\bf 294}, 1495
(2001).
%
\bibitem{Cd2Re2O7-1} M. Hanawa {\em et al.}, Phys. Rev. Lett. {\bf 87}, 187001 (2001).
%
\bibitem{R2Mo2O7} I. K\'{e}zsm\'{a}rki {\em et al.}, Phys. Rev. Lett. {\bf 93}, 266401
(2004).
%
\bibitem{SLpr2ir2o7} S. Nakatsuji {\em et al}., Phys. Rev. Lett.
{\bf 96}, 087204 (2006).
%
\bibitem{AHEpr2ir2o7} Y. Machida, S. Nakatsuji, Y. Maeno, T. Tayama,
T. Sakakibara, and S. Onada, Phys. Rev. Lett. {\bf 98}, 057203
(2007).
%
\bibitem{Fukazawa} H. Fukazawa and Y. Maeno, J. Phys. Soc. Jpn. {\bf 70},
2880 (2001).
%
\bibitem{soda} M. Soda, N. Aito, Y. Kurahashi, Y. Kobayashi, and  M. Sato,
Physica B {\bf 329-333}, 1071 (2003).
%
\bibitem{fukazawa1} H. Fukazawa and Y. Maeno, J. Phys. Soc. Jpn. {\bf 71},
2578 (2002).
%
\bibitem{kbmPRL} For example, K. Maiti, P. Mahadevan, and D.D. Sarma, Phys. Rev.
Lett. {\bf 80}, 2885 (1998); K. Maiti {\em et al.}, Europhys. Lett.,
{\bf 55}, 246 (2001).
%
\bibitem{kartik} We have performed magnetization ($M$) studies on
our material. $M$ exhibits a weak hysteresis at 30 K (i.e., in the
magnetically ordered state) around zero magnetic field ($H$) and
varies rather linearly with $H$ at higher fields, without any
evidence for saturation.  The value of $M$, even at $H$ as high as
120 kOe at 30 K, is negligibly small (about 0.02 $\mu_B$ per formula
unit). On the basis of our results, we infer that there is a
significant antiferromagnetic component, in addition to a
ferromagnetic component.
%
\bibitem{wien} P. Blaha, K. Schwarz, G. K. H. Madsen, D. Kvasnicka,
and J. Luitz, {\scriptsize WIEN2K}, An Augmented Plane Wave + Local
Orbitals Program for Calculating Crystal Properties (Karlheinz
Schwarz, Technical Universität Wien, Austria, 2001), ISBN 3-9501031-
1-2.
%
\bibitem{ddPRL} D.D. Sarma {\em et al.}, Phys. Rev. Lett. {\bf 75}, 1126 (1995).
%
\bibitem{kbmPRB} K. Maiti {\em et al.}, Phys. Rev. B {\bf 73}, 052508 (2006).
%
\bibitem{DDdisorder} D.D. Sarma {\em et al.}, Phys. Rev. Lett. {\bf 80},
4004 (1998).
%
\bibitem{kobayashi} M. Kobayashi, K. Tanaka, A. Fujimori, S. Ray,
and D. D. Sarma, Phys. Rev. Lett. {\bf 98}, 246401 (2007).
%
\bibitem{altshuler-aronov} P.A. Lee and T.V. Ramakrishnan, Rev.
Mod. Phys. {\bf 57}, 287 (1985); B.L. Altshuler and A.G. Aronov,
Solid State Commun. {\bf 30}, 115 (1979).
%
\bibitem{Correlation} A.F. Efros and B.I. Shklovskii, J. Phys.
C {\bf 8}, L49 (1975); J.G. Massey and M. Lee, Phys. Rev. Lett. {\bf
75}, 4266 (1995).
%
\bibitem{bairo3} K. Maiti, R.S. Singh, V. R. R. Medicherla, S. Rayaprol,
and E.V. Sampathkumaran, Phys. Rev. Lett. {\bf 95}, 016404 (2005).
%
\bibitem{irkhin} V.Yu Irkhin, M.I. Katsnelson, and A.V. Trefilov,
J. Phys.: Condens. Matter {\bf 5}, 8763 (1993); V.Yu Irkhin and M.I.
Katsnelson, J. Phys.: Condens. Matter {\bf 2}, 7151 (1990).
%

\end{thebibliography}
\end{document}